\documentclass[runningheads]{llncs}
\usepackage{graphicx}
\usepackage{amsmath}
\usepackage[utf8]{inputenc} %
\usepackage[T1]{fontenc}    %
\usepackage{hyperref}       %
\usepackage{url}            %
\usepackage{booktabs}       %
\usepackage{amsfonts}       %
\usepackage{nicefrac}       %
\usepackage{microtype}      %
\usepackage{cleveref}       %
\usepackage{lipsum}         %
\usepackage{graphicx}
\usepackage{doi}

\usepackage{subcaption}
\usepackage{microtype}
\usepackage{multirow}
\usepackage{makecell}
\usepackage{caption}
\usepackage{hyperref}
\usepackage{float}
\usepackage{enumitem}

\usepackage{booktabs, multirow} %
\usepackage{soul}%
\usepackage[table]{xcolor} %
\usepackage{changepage,threeparttable} %

\usepackage{booktabs}

\begin{document}
\title{
Augmenting Online Classes with an Attention Tracking Tool May Improve Student Engagement
}
\author{
    Arnab Sen Sharma\inst{1}\href{https://orcid.org/0000-0002-0407-6526}{\includegraphics[scale=0.06]{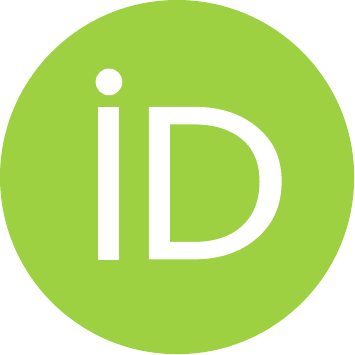}} \and
    Mohammad Ruhul Amin\inst{2}\href{https://orcid.org/0000-0001-6540-3415}{\includegraphics[scale=0.06]{Figures/orcid.pdf}} \and
    Muztaba Fuad\inst{3}
}
\authorrunning{Sen Sharma et al.}
\institute{
    Shahjalal University of Science and Technology, Bangladesh, \email{arnab-cse@sust.edu}\and
    Fordham University, USA, \email{mamin17@fordham.edu} \and
    Winston-Salem State University, USA, \email{fuadmo@wssu.edu}
}
\maketitle              %

\begin{abstract}
Online remote learning has certain advantages, such as higher flexibility and greater inclusiveness. However, a caveat is the teachers' limited ability to monitor student interaction during an online class, especially while teachers are sharing their screens. We have taken feedback from 12 teachers experienced in teaching undergraduate level online classes on the necessity of an attention tracking tool to understand student engagement during an online class. This paper outlines the design of such a monitoring tool that automatically tracks the attentiveness of the whole class by tracking students' gazes on the screen and alerts the teacher when the attention score goes below a certain threshold. We assume the benefits are twofold; 1) teachers will be able to ascertain if the students are attentive or being engaged with the lecture contents and 2) the students will become more attentive in online classes because of this passive monitoring system. In this paper, we present the preliminary design and feasibility of using the proposed tool and discuss its applicability in augmenting online classes. Finally, we surveyed with 31 students asking their opinion on the usability as well as the ethical and privacy concerns of using such a monitoring tool. 

\end{abstract}

\section{Introduction}
Most of the educational activities worldwide had shifted to online/remote learning during the COVID-19 pandemic outbreak to curtail its growth \cite{lall2020covid}, \cite{yu2020analysis}, and \cite{pokhrel2021literature}. Teachers and students in every level of educational institutions (schools, colleges, and universities) have embraced and become adept at using online platforms and tools during this pandemic. Remote learning presents many unique opportunities including higher flexibility and greater inclusiveness \cite{ferri2020online}. For these reasons, this form of education is very likely to continue even in the post-COVID era. Many educational institutions have already introduced a hybrid system, a combination of online and offline classes \cite{adedoyin2020covid}. Class engagement plays an important role in the learning process \cite{webster1993dimensionality}, but it is difficult for instructors to monitor student interaction during an online class \cite{zhang2020wandering}. \\

To understand the problem with student engagement in online classes, we organized a round-table discussion on \textit{"Ensuring Student Engagement in Online Classes"} with 12 university teachers who have taught online undergraduate courses during the lockdown period of COVID-19 pandemic. Among the participants, 8 were male and 4 were female. We requested them to share the challenges they have faced with student engagement while teaching online classes. We summarize the concerns raised in our discussion below.

\begin{itemize}
    \item It is somewhat impossible to monitor all the students during an online class conducted through \textit{Zoom}, \textit{Google Meet}, or any other platforms they have tried. Especially, while the teacher is sharing the screen and presenting slides. On the other hand, many of the students feel reluctant to share their video feedback during an online class due to privacy issues or low internet bandwidth. 
    \item Handful of the teachers informed their concerns that they have no way to understand whether students are listening to the lecture or not. The teachers mentioned they had to ask the class many times - “\textit{Are you guys with me?}” or “\textit{Does this make sense?}” - in order to get verbal feedback. However, usually the responses come from high-performing students while the majority remain silent. 
    This is why all of the teachers we discussed with corroborate the importance of having some way of understanding the level of student engagement. In other words, they want to know how many of the students are listening to the lecture actively.
    \item  Many of the teachers are also worried that their students might be taking online classes less seriously. They suspect that students might be engaging in other activities like; browsing social media or different websites while class is in progress. The teachers feel that since there is no way of making direct eye contact with the students of online classes, a monitoring tool might help make the students more attentive. 
\end{itemize}

\begin{figure}
\centering
    \includegraphics[totalheight=5.5cm]{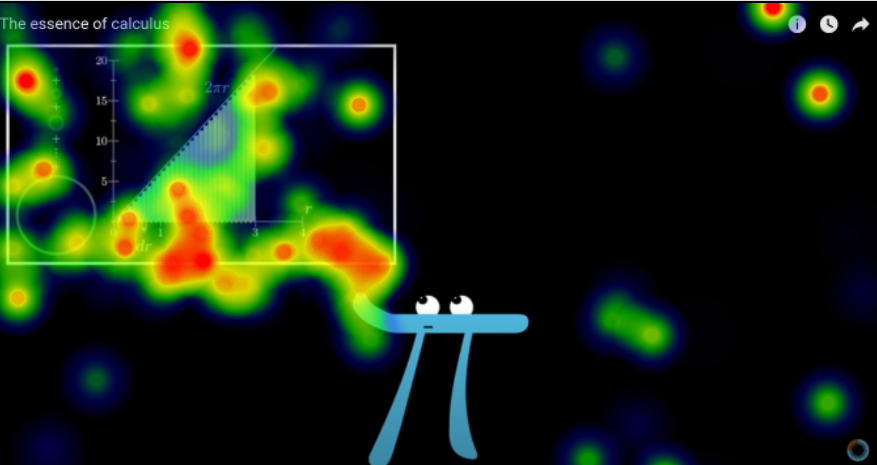}
    \caption{A \textit{YouTube} video named “The essence of calculus” by \textit{3Blue1Brown} was shown to 3 students. The video was maximized to full-screen and students’ gaze was recorded. The heatmap overlay on the figure represents students’ gazes from 7:06 to 7:12 when the narrator focuses on the animation depicted in the top-left corner. The undergrad students were requested to try their best to maintain their full concentration while they watch the video clip. And, as expected, their gazes co-align (represented by the high concentration of gazes in the focused animation}
    \label{fig: 3b1r}
\end{figure}

To address the above concerns, we present a tool that can help an instructor teaching an online class get better insights into how well the class is following the lecture. 
The main idea behind the tool is that at any particular moment during an online class, usually the instructor attracts students’ attention to a particular region of the screen. If the students are indeed attentive to the class, their gaze patterns should align to some extent (\textbf{Figure \ref{fig: 3b1r}}). The tool has been designed to run in the background during an online class and collect gaze information of students via an webcam. To eschew ethical concerns, this tool does not collect video feedback from students, rather it collects students gaze on screen in the form of \texttt{x,y} coordinates. The tool will measure the density of the students' collective gaze in a particular screen area to rate the level of attention of the whole class. The instructor can utilize this score to gauge the engagement of students with the class.\\

In this study, we discuss the design and feasibility of this monitoring tool using existing gaze-tracking technology compatible with low-cost webcams. We have collected gaze-stream data from 31 students using a prototype and discuss the performance of the gaze-tracking module. We hypothesize that certain patterns emerge in the aggregated gaze distribution while students are paying attention to a particular focus object (i.e., the item on screen the teacher is explaining) and verify our hypothesis using randomization test. Furthermore, we discuss the considerations and challenges to calculate an \textit{attention} score for the whole class. Finally, we conduct individual interviews with students to get their perspective on the ethical and privacy concerns of using such a tool.

\section{Related Works}
Several studies have found positive correlations between visual attention and performance \cite{yantis1990abrupt}, \cite{prinzmetal1986does}. Finn \cite{finn1989withdrawing} argued strong correlation between students' academic performance and class attention. Hutt et al. \cite{hutt2017gaze} and Wammes et al. \cite{wammes2016mind} discussed the importance of monitoring students' attention state during a class. E Campbell \cite{emma2014} tried to establish a link between classroom engagement and the gaze pattern of primary-school aged students and found that students were more inclined to direct their gaze into their own work or the works of their classmates during the classroom discussion instead of fixating their gaze on teachers instructions. However, in our case, we are interested in measuring the attention of university-level students in online classes. University level theory classes primarily consist of lectures and students usually have less classroom activity apart from listening to and engaging with instructor's lecture.\\
    
Sharma et al \cite{sharma2014me} introduced a gaze-based student-teacher co-attention score named \texttt{with-me-ness} which tries to answer a question from teachers' perspective - "\textit{How much are the students with me?}". They conducted a study by showing a MOOC video lecture to 40 participants and found positive correlation between the \texttt{with-me-ness} scores and scores achieved by students in a test conducted after showing the video lecture. They measure this \texttt{with-me-ness} value at two levels: \textit{Perceptual} (if a student is looking at the items being referred by teacher's deictic acts  -- highlighting, annotating using digital pen, etc) and \textit{Conceptual} (do students' look at the item the teacher is verbally referring to). Calculating this \texttt{with-me-ness} score requires extensive preprocessing steps that will be difficult or impossible to perform before every online class. In this study, we are interested in a system that works in real-time and does not require extensive preprocessing steps. Also the \textit{Conceptual} part of the \texttt{with-me-ness} score can be infeasible to measure in real-time as the system will need to have extensive expertise in the topic being instructed as well as speech processing, natural language processing, and others. Our approach of calculating attention score rather depends on the desnsity of collective gazing points on the screen. We argue that if the students are following the class their gaze patterns should follow regions of screens the teacher is attracting their attention and thus the gaze patterns should automatically align among themselves. An obvious limitation of our system is that it will fail to  function when there is only one student in the class, which is rarely the case.\\

Srivastava et al \cite{srivastava2021you} extends Sharma et al's \cite{sharma2014me} work by introducing a matric \texttt{with-me-ness direction}, which tries to measure the video-watching attention patterns between a student's gaze and instructor's dialogue. They show that students gaze patterns can vary with their prior knowledge of the topic. And similar to \texttt{with-me-ness}, \texttt{with-me-ness direction} requires extensive preprocessing steps as well. \\

De Carolis et al. \cite{de2019engaged} has classified student engagement in a 4-point scale using Long Short-Term Memory (LSTMs) based on facial expressions, head movement, and gaze behavior. They have used the \textit{OpenFace 2.0} toolkit to extract the eye-gaze estimation, head-pose, facial features from the video feed of a student. Yang et al. \cite{yang2018student} and Rahul et al. \cite{rahul2020real} propose similar approaches for tracking attention during an online class using facial expressions, posture, and gaze pattern. Unlike our setting, all of the mentioned approaches expect video feed from students during an online class. Whereas, we only use a light-weight tool to track students gaze on the client-side.\\

According to the \textit{eye-mind hypothesis} by Just et al. \cite{just1980theory}, we learn and process the information that we visually attend to. However, it is difficult to establish that visual attention transfers to learning activity or greater visual attention will result in better academic performance. But, in an online class setting, it is one of the very few information that can be exploited to \textit{measure} student engagement in a class. In this study, we introduce a tool capable of following students gaze patterns during an online class, establish the usefulness of such tool by conducting interviews with university level teachers and students, and try to address possible ethical concerns raised by students while designing this system.

\section{Design Considerations}
A high-level overview of the tool is depicted in \textbf{Figure \ref{fig: overview}} below. An web application will interact with the webcam to collect gaze information during an online class (\textbf{Figure \ref{fig: overview}a}). The gaze-stream  in the form of 2D cartesian coordinates (\texttt{x,y})  will be sent to a server. At the server side, this gaze patterns will be aggregated to calculate an attention score for the whole class (\textbf{Figure \ref{fig: overview}b}). The instructor will be alerted when the attention score goes below a predefined threshold. While designing this tool we took several considerations into account discussed below. 

\begin{figure}
\centering
    \includegraphics[totalheight=5.5cm]{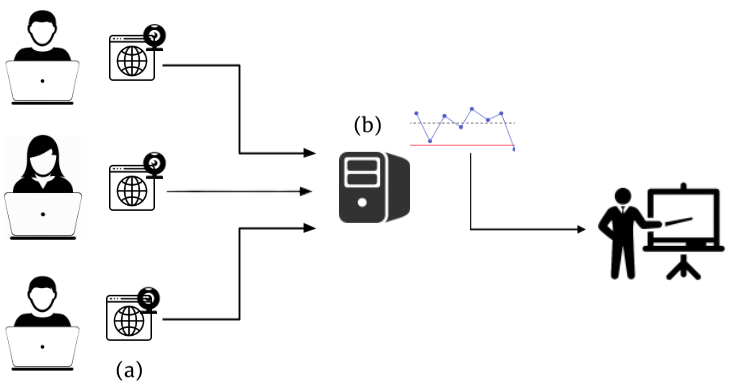}
    \caption{Overview of the attention tracking tool.}
    \label{fig: overview}
\end{figure}

\begin{itemize}
    \item \textbf{Portability:} To ensure greater inclusiveness by not binding the students and teachers into specific operating systems or devices, we implemented this tool as a web application. This tool can be used with a modern browser on any device with a webcam running any operating system. The web application will run in the background while the online class is running on another platform (\textit{zoom}, \textit{meet}, or another tab of the browser). The application will download an eye-tracking module on clients’ machines that runs in the browser sandbox.
    
    \item \textbf{Light-weight:} We cannot expect all the students to have high-end devices. So, we ensured that the tool is not too resource-hungry and does not require special processing units such as GPUs or TPUs to run. In practice, we adapt \textit{WebGazer}\footnote{\url{https://webgazer.cs.brown.edu/}}, an open-source eye-tracking library by Papoutsaki et al \cite{papoutsaki2016webgazer}, which is implemented in JavaScript and it can be integrated into a web application. At the core, this library uses an eye-detection library to detect the pupils and then uses their location to linearly estimate the location on the screen where a user’s current gaze is at. There is a brief calibration phase required before the tool can be used. 
    
    \item \textbf{Ethical Concerns:} This tool does not expect video feedback from students. And, the attention score will be calculated for the whole class analysing the gaze-pattern of every student present in the class. This score will be used to alert the instructor only when the score goes below a certain threshold. So, the teacher will have no way of identifying the gaze pattern of an individual student.
    The main purpose of this tool is to make both students and the teacher self-conscious in order to ensure better class engagement. We expect that students will be more diligent during online classes because of this monitoring tool. Also, the students will be more accepting towards this monitoring tool as it has relatively less effect on their privacy. The teacher can use the attention score to understand when the class has become too difficult/tedious and upgrade their teaching styles and lecture contents.
\end{itemize}

In the following sections we discuss the feasibility of eye-tracking technologies with low-cost webcams to estimate gaze patterns which is an essential part of our tool. We set up a study with 31 undergraduate students to ascertain this tool’s feasibility.

\section{Data Collection}
We arranged individual online meetings with each of the 31 students via \textit{Zoom}, a popular platform of choice for online classes. Each of the participants was instructed beforehand to participate from a laptop or desktop with a webcam. We encouraged the participants to join in the sessions in the setting while they usually participate in online classes, which means: diverse backgrounds and different ambiance setups. Also, students were allowed to wear prescription glasses if they prefer to do so. Each session took around 20-25 minutes and each of the participants was paid 100 takas ($\sim$ 1.2 USD) in the form of mobile balance recharge for their contribution to this study.

\begin{table}[]
    \centering
    \begin{tabular}{|c|c|}
        \hline
        \textbf{Male}   & \textbf{Female}   \\
        \hline
            27          &     4             \\
        \hline
    \end{tabular}
    \quad
    \quad
    \begin{tabular}{|c|c|}
        \hline
        \textbf{Low}   & \textbf{High}\\   
        \textit{(0.3MP)} & \textit{($\ge$ 0.9)}\\
        \hline
            12          &     19             \\
        \hline
    \end{tabular}
    \quad
    \quad
    \begin{tabular}{|c|c|}
        \hline
        \textbf{With}   & \textbf{Without} \\  
        \textbf{Glasses} & \textbf{Glasses}\\
        \hline
            11          &     20             \\
        \hline
    \end{tabular}
    \vspace{.2cm}
    \caption{Participants distribution in gender, their webcam resolution, and whether they wear prescription glasses or not. Total number of participants is \textbf{31}}
    \label{tab:participant_distribution}
\end{table}

During the calibration phase, the participants were instructed to directly look at and click 3 times on some focus points. These focus points appear one by one, a total of 20 times at predefined locations on the screen in a random sequence. The entire calibration phase takes about 35-40 seconds to complete. Although this calibration phase is required only once per device, we have found that performing calibration before each class results in improved gaze estimation performance.

\begin{figure*}[t!]
\hspace*{-1in}
    \centering
    \begin{subfigure}[t]{0.5\textwidth}
        \centering
        \includegraphics[width=1.35\textwidth]{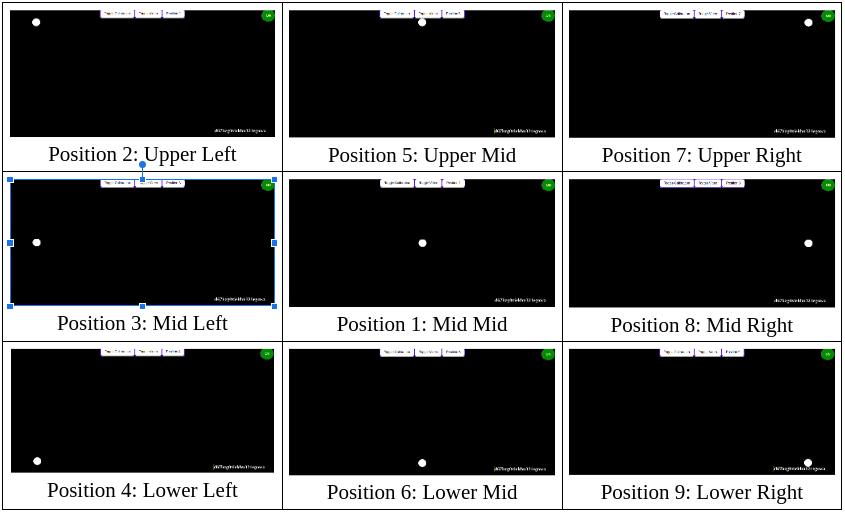}
        \caption{Focus points for different regions}
        \label{fig: screen_regions}
    \end{subfigure}%
\hspace{0.9in}
    \begin{subfigure}[t]{0.5\textwidth}
        \centering
        \includegraphics[width=1.35\textwidth]{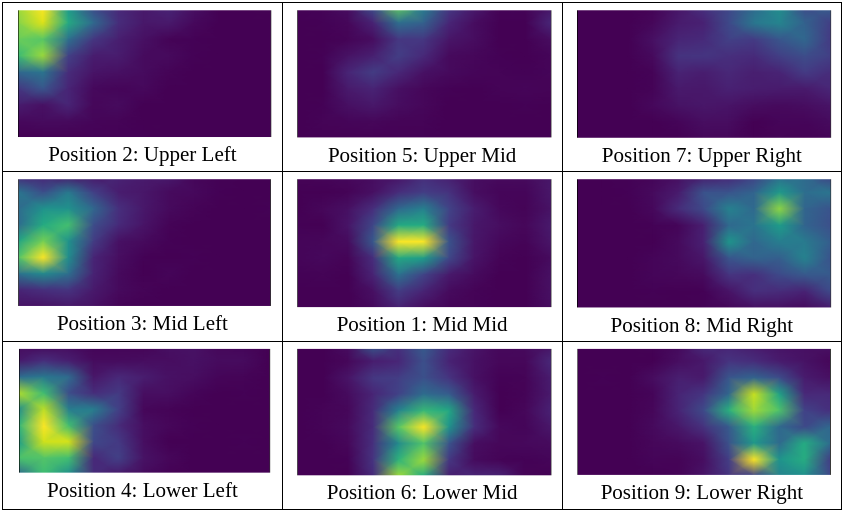}
        \caption{Region-wise heatmaps of collected gaze patterns (for 31 students)}
        \label{fig: heatmaps}
    \end{subfigure}
    \caption{Calibrating (a), heat-map of gazing (b)}
    \label{fig: regions_and_heatmaps}
\end{figure*}

We divided the screen into 9 regions. And we instructed the participants to focus their gaze on one of the regions for $\sim10$ seconds while we record their gaze estimated by \textit{WebGazer}’s eye-tracking module. To help the participants focus their gaze on a specific region of the screen, we show a focus point (\textbf{Figure \ref{fig: screen_regions}}). 
All of the gaze points (\texttt{x, y} coordinates) received from students were divided by their screen dimensions (\texttt{x/screen-width} and \texttt{y/screen-height}) in order to normalize for different screen sizes and resolutions.

\section{Gaze Estimation Result and Error Analysis}
\subsection{Emerged Gaze Patterns}
The region-wise collected heatmaps of combined gaze-streams of 31 students are depicted on \textbf{Figure \ref{fig: heatmaps}}. We can observe that almost in all cases the estimated gazes are concentrated in close proximity to respective focus points (\textbf{Figure \ref{fig: screen_regions}}). This attests to this tool's reliability at predicting users' gaze on screen. We conduct randomization tests to verify that the aggregated gaze-stream for a specific region is different and separable from uniformly distributed 2D points.
For this purpose, we coin a term \textit{cohesiveness} which we define as the MSE loss for a gaze distribution with respect to its \textit{centroid}.

\begin{equation}
    cohesiveness = \frac{1}{N}\sum\limits_{i=1}^N((x_{centroid} - x_i)^2 + (y_{centroid} - y_i)^2)
\end{equation}
where $N$ is the number of points in a gaze distribution $(x_i, y_i)$, \\and $(x_{centroid}, y_{centroid})$ represents the \textit{centroid} of the distribution\\

To perform randomization test that gaze distribution is different from any random distribution of 2D points, we state the \textit{null} hypothesis and \textit{alternative} hypothesis below.

\vspace{.3cm}
Let, 
\vspace{-.23cm}
\begin{itemize}
    \item \texttt{Random-Focus Diff} is the \textit{cohesiveness} difference between a gaze distribution focused at a specific region and a random distribution of 2D points.
    \item \texttt{Random-Random Diff} is the \textit{cohesiveness} difference between two random uniform 2D distributions.
\end{itemize}

Then,
\vspace{-.23cm}
\begin{itemize}
    \item $H_0$: Focused gaze-distribution is \textit{not} different from a random uniform 2D distribution. Meaning, the values \texttt{Random-Focus Diff} and \texttt{Random-Random Diff} are similar. 
    \item $H_1$: Focused gaze-distribution is \textit{significantly} different from a random uniform 2D distribution. That means, the values \texttt{Random-Focus Diff} and \texttt{Random-Random Diff} are separable with a clear margin. 
\end{itemize}

\begin{figure}
\centering
    \includegraphics[totalheight=6cm]{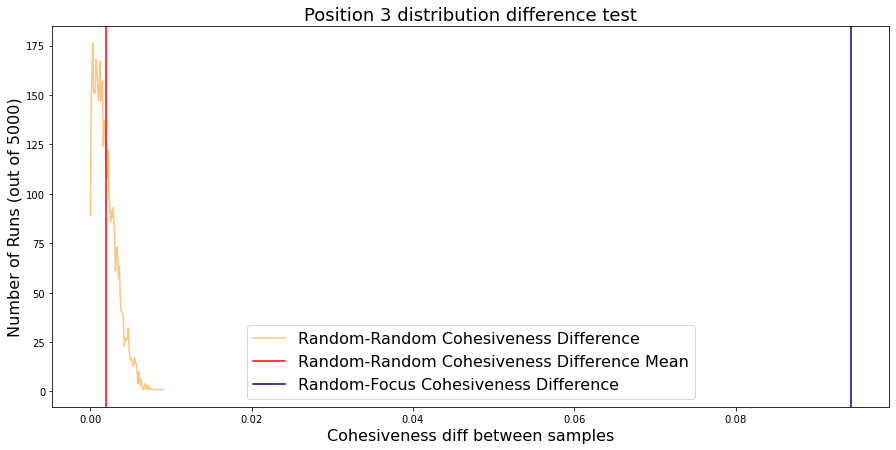}
    \caption{Randomization Test Results.}
    \label{fig: rand_test}
\end{figure}

\vspace{.3cm}
Our randomization test consists of the following steps:

\begin{enumerate} [label=Step \arabic*:, leftmargin=*]
    \item For each of the nine regions (\textbf{Figure \ref{fig: screen_regions}}), we take the gaze distribution $(x_i, y_i)$. Then, we take a random sample of $(x, y)$ co-ordinates from a uniformly distributed 2D points. To compare these two distributions, we calculate the difference between their \textit{cohesiveness}.
    For example, for a specific region, we have the \texttt{focused} gaze pattern and the random uniform distribution depicted in \textbf{Figure \ref{fig: rand_test_pos_3}}. Using \textbf{Equation 1}, we calculate the \textit{cohesiveness} values for respective distributions shown below. To test the hypothesis, we will use the \textit{cohesiveness} difference of the actual gaze and random uniform distribution (\texttt{Random-Focus Diff}).   

    \begin{itemize}
    \centering
        \item[] \texttt{Random Cohesiveness :  0.17023544468407120}
        \item[] \texttt{Focus Cohesiveness ~:  0.07365995468797122}
        \item[] \texttt{Random-Focus Diff ~~:  0.09657548999609998}
    \end{itemize}
    
    \begin{figure}[h!]
    \centering
        \includegraphics[totalheight=4cm]{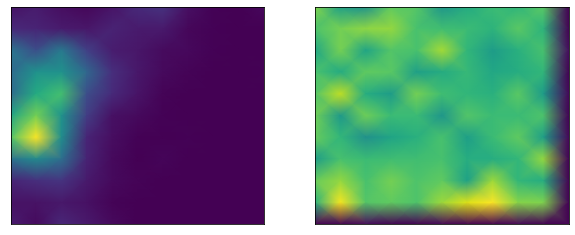}
        \caption{Actual Gaze distribution vs Random uniform distribution}
        \label{fig: rand_test_pos_3}
    \end{figure}

    \item To perform the randomization test, we compute the difference of \textit{cohesiveness} between two random samples of size 5000. We measure the \textit{cohesiveness} difference between these two new distributions (\texttt{Random-Random Diff}). And compare the \texttt{Random-Random Diff} with \texttt{Random-Focus Diff}.

    \item We run this simulation for 5000 times and plot the frequency of \textit{cohesiveness} differences in \textbf{Figure \ref{fig: rand_test}} (the orange-colored line chart).
\end{enumerate}

In \textbf{Figure \ref{fig: rand_test}}, we can observe clear distinction between \texttt{Random-Random Diff} with \texttt{Random-Focus Diff}. In the plot, we can find the \texttt{Random-Random Diff} as the orange line-plot, where we observe that the smallest \textit{cohesiveness} difference has the highest frequency of occurrence and it goes down as the \textit{cohesiveness} difference goes higher. The red vertical line represents the average of the \texttt{Random-Random Diff} values. The blue vertical line which is further apart on the rightmost side, represents the \texttt{Random-Focus Diff}. As we perform z-test to find if the blue vertical line is similar to the distribution in orange color, we find that it is rather significantly different at $p << 0.05$. Thus, we reject the \textit{null} hypothesis and accept the \textit{alternative} one. Similar trends were observed for all the 9 focus regions of \textbf{Figure \ref{fig: screen_regions}}. 

Thus, we can decisively conclude that gaze patterns emerge when multiple participants focused their gaze on a specific region and \textit{Webgazer} is capable of capturing this pattern.

\subsection{Error Analysis of the Gaze-Tracking Module}
\begin{figure}[t]
    \centering
    \begin{subfigure}[t]{0.5\textwidth}
        \centering
        \includegraphics[height=3in]{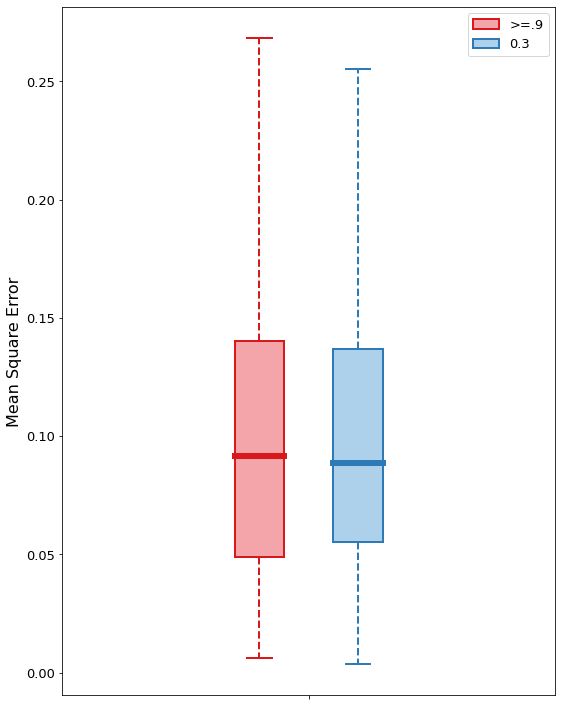}
        \caption{Performance in different camera resolutions}
        \label{fig: comp_res}
    \end{subfigure}%
    \begin{subfigure}[t]{0.5\textwidth}
        \centering
        \includegraphics[height=3in]{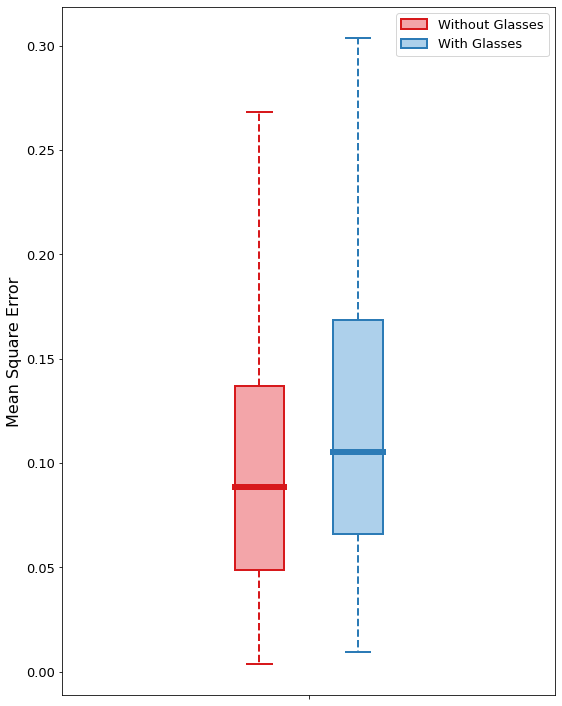}
        \caption{Performance with vs without glasses}
        \label{fig: comp_glass}
    \end{subfigure}
    \caption{Mean square error of our presented tool in different camera resolutions (left) and whether the student was wearing prescription glasses or not (right). Our tool achieves a satisfactory performance with \texttt{MSE} $\le$ 0.1 at predicting the gaze focus according to our collected data. Note that, all of the gaze points (\texttt{x, y} coordinates) were divided by the screen dimensions (\texttt{x/screen-width} and \texttt{y/screen-height}) in order to normalize for different screen sizes and resolutions.}
    \label{fig: performace_comparison}
\end{figure}

We calculated the mean square error of the gaze tracking module in different settings and find that the eye-tracking module yields satisfactory performance on average (\textbf{Figure \ref{fig: performace_comparison}}). Our gaze tracking tool archives a mean-square-error of less than 0.1 on average. Only when students are wearing prescription glasses the MSE slightly exceeds 0.1.

\section{Considerations and Challenges to Measure Attention Score from Gaze Patterns}
We have yet to finalize the procedure to calculate the attention score from a gaze distribution. However, we have been experimenting with a few ideas. Till now, we have found that \textbf{DBScan}, a clustering algorithm is showing some promising results.  We have used the \textbf{DBScan} implementation of \texttt{scikit-learn} library by Pedregosa et al. \cite{scikit-learn} which expects two hyperparameters: 

\begin{itemize}
    \item \textit{Epsilon}, in short \textbf{\texttt{eps}}: The maximum distance allowed between two points to be considered belonging to the same cluster. That means a point \textit{\textbf{p}} will belong in a cluster \textbf{C} if and only if there is atleast one point \textit{\textbf{q}} in \textbf{C} such that the distance between \textit{\textbf{p}} and \textit{\textbf{q}} is less than or equal \textbf{\texttt{eps}}.
    \item \textbf{\texttt{min\_samples}}: The minimum number of samples in a cluster for a point to be considered as a core point. Effectively, there is atleast \textbf{\texttt{min\_sample}} number of neighbouring points required to form a cluster. 
\end{itemize}

\begin{figure}
\centering
    \includegraphics[totalheight=7cm]{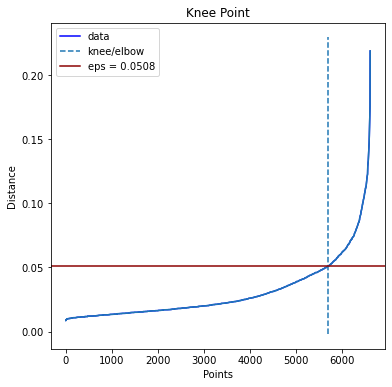}
    \caption{Tuning the \textbf{\texttt{eps}} value for \textbf{Position 1}.}
    \label{fig: tuning_eps}
\end{figure}

During our experiments, we set the \textbf{\texttt{min\_samples}} value to $100$. This number has been established based on the observations during our experiments. We tune the \textbf{\texttt{eps}} value \textit{dynamically} while analysing a collection of gaze patterns. The idea\footnote{How to Use DBSCAN Effectively, link: \url{https://towardsdatascience.com/how-to-use-dbscan-effectively-ed212c02e62}} is that, we can use nearest neighbours to reach a fair estimation of an \textit{optimum} \textbf{\texttt{eps}} value. For each point, we find its distance from its $\lfloor min\_samples/3 \rfloor ^{th}$ closest neighbour. Sorting these distance values give us an \textit{exponential} like curve depicted at \textbf{Figure \ref{fig: tuning_eps}}. The \textit{optimum} \textbf{\texttt{eps}} value would be the \textbf{\texttt{y}} value at the \textbf{elbow} of the curve. There is a sharp rise after the \textbf{elbow}, which means some points are really sparse and they can be discarded as noise. We use the \textbf{\texttt{kneedle}} library by Ville Satopaa et al \cite{satopaa2011finding} to find the \textbf{elbow} point. 

\begin{figure}
\centering
    \hspace*{-1.3in}
    \includegraphics[totalheight=15cm]{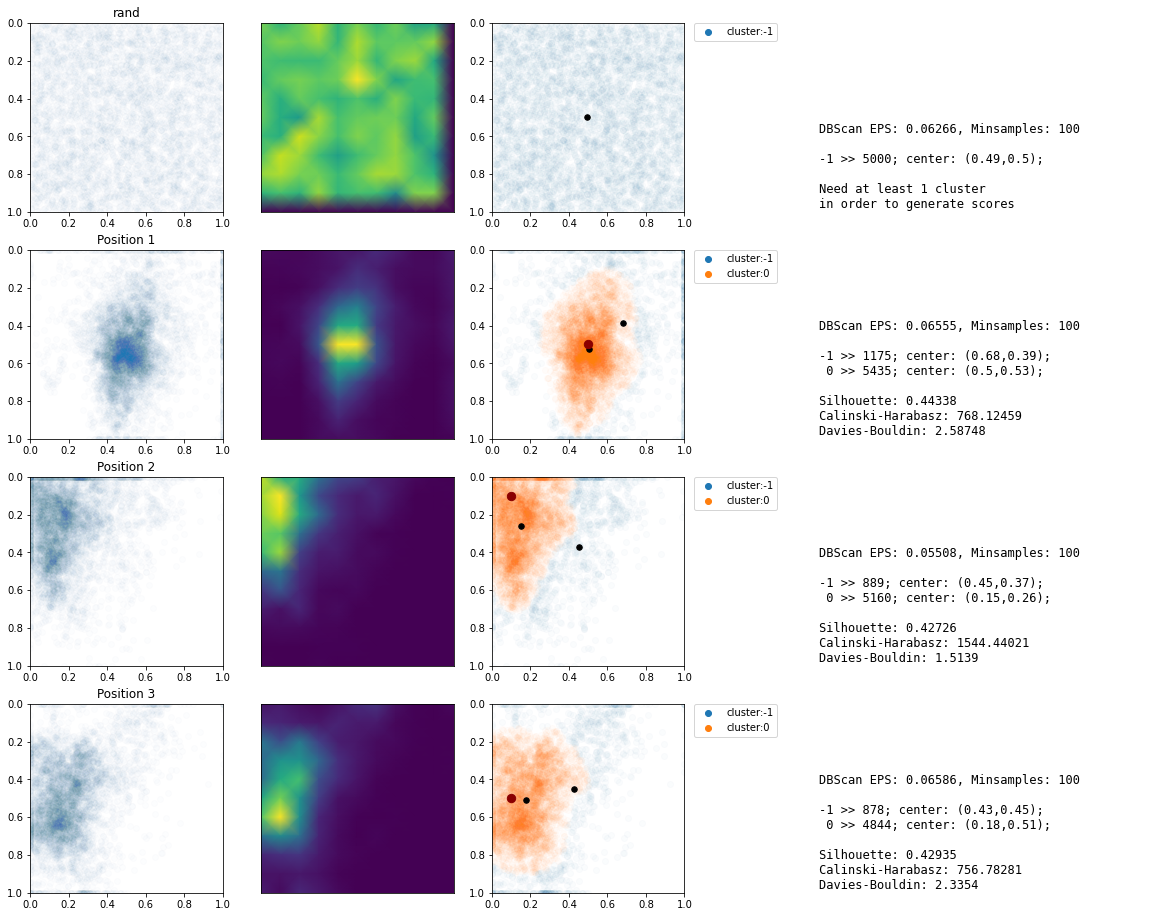}
    \caption{\textbf{DBScan} performance on aggregated gaze-streams from a single focus region}
    \label{fig: dbscan_same}
\end{figure}

\begin{figure}
\centering
    \hspace*{-1.3in}
    \includegraphics[totalheight=12cm]{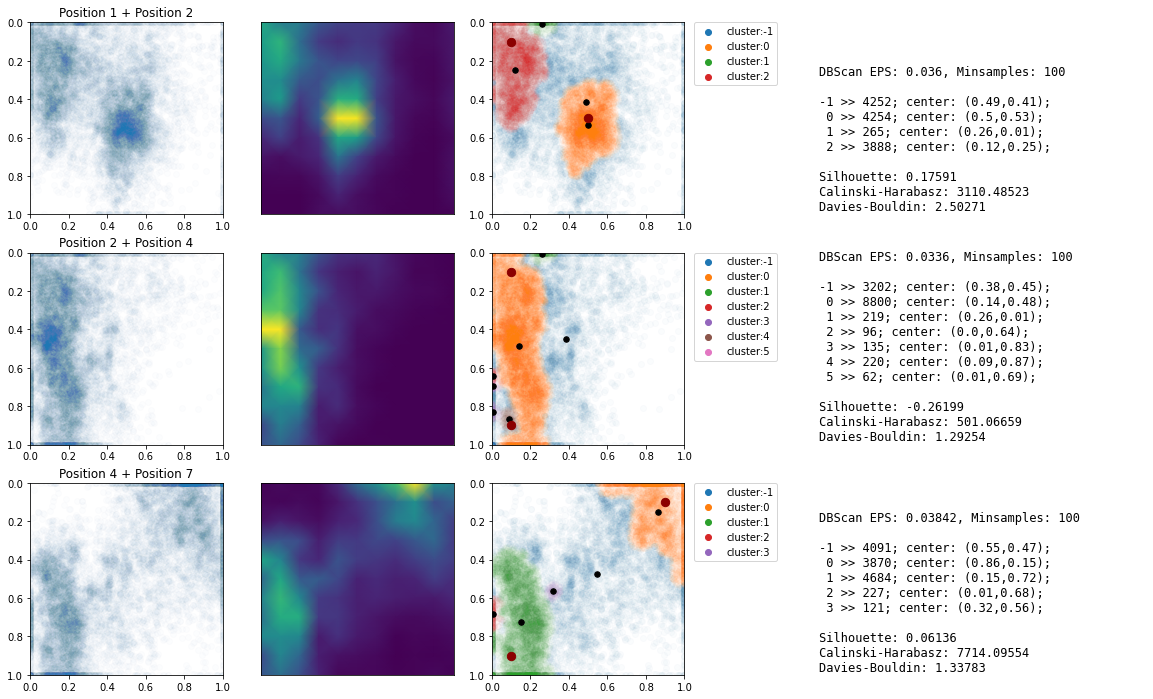}
    \caption{\textbf{DBScan} performance on aggregated gaze-streams from two different focus regions}
    \label{fig: dbscan_different}
\end{figure}

The performance of \textbf{DBScan} with tuning is depicted on \textbf{Figures \ref{fig: dbscan_same}} and \textbf{\ref{fig: dbscan_different}}. \textbf{Figure \ref{fig: dbscan_same}} compares the performance of \textbf{DBScan} between a random distribution and the gaze distribution collected for focused regions. And \textbf{Figure \ref{fig: dbscan_different}} shows the results of \textbf{DBScan} when the gaze distributions of two different focused regions get mixed. We can observe clear distinction in the \textbf{DBScan} results and we are considering how this difference can be utilized to calculate an attention score. 

\begin{figure}
\centering
    \hspace*{-1.3in}
    \includegraphics[totalheight=12cm]{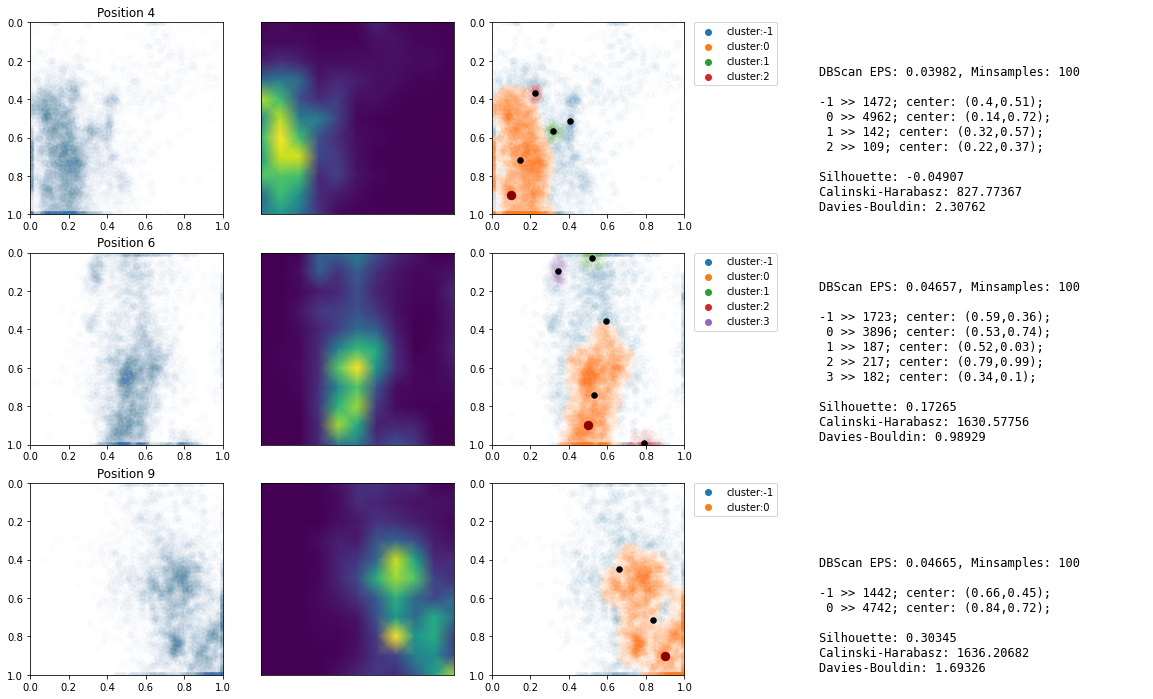}
    \caption{\textbf{DBScan} \textit{bad} performance on aggregated gaze-streams from a single focus region}
    \label{fig: dbscan_same_bad}
\end{figure}

There are some certain caveats, though. To reach a reliable attention score calculation method we need consistent results. But, in our collected data we observe some relatively bad clustering results of \textbf{DBScan}. In \textbf{Figure \ref{fig: dbscan_same_bad}}, we observe that for positions 4 and 6, \textbf{DBScan} identifies 3 and 4 clusters respectively. We have used the \textit{WebGazer} library for gaze-tracking, which still has some room for improvements. During the calibration phase of our data collection the participants complained that the tracking module was not able to follow their gazes properly in the bottom regions of the screen, especially while the participants were wearing prescription glasses. There is a pupil-detection module at the core of the \textit{WebGazer} library. We think that when a user is wearing glasses this pupil-detection module renders low performance. Further investigation is required in this regard.

\section{Students Perspective of the Monitoring Tool}
Following the data collection session, we conducted short interviews with the participants to ascertain the usability of this monitoring tool in a real-world online class setting. The findings from our discussions are summarized below.

\begin{itemize}
    \item \textbf{Ease of use}: All of the participants found the tool to be easy to use. The students appreciated that it does not require them to install any new software as the tool works with a browser. Furthermore, the calibration phase takes only around 20-25 seconds. However, students who prefer to wear prescription glasses faced some issues during the calibration phase. They complained that the \textit{Webgazer} tool could not track their gaze properly at the bottom layer of the screen (positions 4, 6 and 9).
    
    \item \textbf{Low resource and bandwidth consumption}: Students appreciated the fact that our tool consumes very low compute resource and internet bandwidth. Our tool was able to perform perfectly in relatively low-end devices (with core i3 processors and 2GB of RAM) without sacrificing the performance while the interview was in-progress via \textit{Zoom}. The students were also content that this monitoring tool does not add much to the bandwidth payload. 
    
    \item \textbf{Ethical and privacy concerns}: The participants were assured that only their gaze-stream data will be uploaded to a server where it will be aggregated into an attention measurement function. Thus, no one will have access to the video feed and the gaze data of an individual student. While 18 out of 31 participants were reluctant to share their video feed, only 2 participants raised concerns with sharing their gaze information during an online classes.
    
    \item \textbf{Efficacy of a passive monitoring tool}: We asked the students some additional questions regarding their attention span during an online classes: \textit{why they lose their attention}, and \textit{how likely this tool might help regarding this issue}. Some of the students blamed notification sounds of their smartphones from social media and messaging applications. In an online class with video feedback turned off, due to the lack of supervision some students (12 out of 31) said that they are often inclined to browse their social media sites when their attention gets interrupted by a notification or they find the lesson too difficult or tedious. They feel that as this tool might help the teacher understand when the lecture content is getting too difficult or monotonous for the students to follow, the teacher might be inclined to invest more time to explain the topic slowly, in a more understandable manner to reach most of the students.
    
    \item \textbf{Concerns raised by students}: Though the tool will not enable a teacher to monitor the \textit{attentiveness} of an individual student some students raised concerns that the teacher might have a negative impression for the classes that have a low overall attention score. Some students raised concerns that this negative impression might have some impact on the overall grading for the course.
\end{itemize}

\section{Conclusion}

In this paper, we discuss the design and feasibility of a new monitoring tool that can augment online classes and make them more engaging. This tool will require only gaze-stream information in the form of \texttt{x, y} coordinates to function. We ascertain the necessity and feasibility of this tool in real-online classes consulting with both students and teachers. An obvious shortcoming of our tool is that it will fail to function properly if there is only one student attending the class. However, that is usually not the case. We discuss the considerations and challenges to finalize a method for calculating a reliable attention score by aggregating the continuous gaze-stream data from students during an online class. This work is very much in the preliminary phase. In this paper, we only present the results based on the data we have collected during our individual interview sessions with 31 students. As of now, we have not conducted any real online classes. To verify how our tool performs at deducing different levels of student engagement we will have to check how the attention scores correlate with student engagement of real classes. We have plans to take at least 10 online classes with 30+ students. The classes will be designed to allow a variable level of student engagement and each student will be directly monitored by a proctor to ascertain the level of his/her engagement during the course of each class. The data collected during these classes will serve as the gold standard for establishing the attention measurement methodology. 

We pledge to make the tool open-source. Also, all our collected data and analysis code-base will be made publicly available. Our work-in-progress codebase can be accessed with the links below.
{\color{blue}
\begin{itemize}
    \item[] \href{https://github.com/arnab-api/AttentionChecker--Student}{Student-Side (the gaze-tracking module)}
    \item[] \href{https://github.com/arnab-api/AttentionChecker--Instructor}{Instructor-side (attention measurement module)}
\end{itemize}
}

\bibliographystyle{unsrt} 
\bibliography{references}

\end{document}